\documentclass[conference]{IEEEtran}
\IEEEoverridecommandlockouts
\usepackage{cite}
\usepackage{amsmath,amssymb,amsfonts}
\usepackage{algorithmic}
\usepackage{graphicx}
\usepackage{textcomp}
\usepackage{subcaption}
\usepackage{xcolor}
\usepackage{float}
\usepackage{multirow}
\usepackage{bigstrut}
\usepackage{makecell}
\def\BibTeX{{\rm B\kern-.05em{\sc i\kern-.025em b}\kern-.08em
    T\kern-.1667em\lower.7ex\hbox{E}\kern-.125emX}}
\begin{document}

\title{Simple Yet Surprisingly Effective Training Strategies for LSTMs in Sensor-Based Human Activity Recognition}

\newcommand{\thomas}[1]{\textcolor{blue}{[#1 -- TP]}}

\author{\IEEEauthorblockN{Shuai Shao}
\IEEEauthorblockA{\textit{Department of Computer Science} \\
\textit{University of Warwick}\\
Coventry, UK \\
shuai.shao.1@warwick.ac.uk}
\and
\IEEEauthorblockN{Yu Guan}
\IEEEauthorblockA{\textit{Department of Computer Science} \\
\textit{University of Warwick}\\
Coventry, UK \\
yu.guan@warwick.ac.uk}
\and
\IEEEauthorblockN{Xin Guan}
\IEEEauthorblockA{\textit{School of Computing} \\
\textit{Newcastle University}\\
Newcastle upon Tyne, UK \\
xin.guan@newcastle.ac.uk}
\and
\IEEEauthorblockN{ \hspace{100pt}Paolo Missier}
\IEEEauthorblockA{\textit{ \hspace{100pt}  School of Computing} \\
\textit{ \hspace{100pt} Newcastle University}\\
\hspace{100pt} Newcastle upon Tyne, UK \\
\hspace{100pt} paolo.missier@newcastle.ac.uk}
\and
 \IEEEauthorblockN{Thomas Ploetz}
\IEEEauthorblockA{\textit{School of Interactive Computing} \\
\textit{Georgia Institute of Technology}\\
Atlanta, USA \\
thomas.ploetz@gatech.edu}

}

\maketitle

\begin{abstract}
Human Activity Recognition (HAR) is one of the core research areas in mobile and wearable computing. 
With the application of deep learning (DL) techniques such as CNN, recognizing periodic/static activities (e.g, walking, lying, cycling, etc.) has become a well-studied problem.
What remains a major challenge though is the sporadic activity recognition (SAR) problem, where activities of interest tend to be non-periodic, and occur less frequently when compared with the often large amount of irrelevant ``background'' activities. 
Recent works suggested that sequential DL models (such as LSTMs) have great potential for modeling non-periodic behaviours, and in this paper we studied some LSTM training strategies for SAR. 
Specifically, we proposed two simple yet effective LSTM variants, namely delay model and inverse model, for two SAR scenarios (with and without time-critical requirement).
For time-critical SAR, the delay model can effectively exploit pre-defined delay intervals (within tolerance) in form of contextual information for improved performance. 
For regular SAR task, the second proposed,  inverse model can learn patterns from the time-series in an inverse manner, which can be complementary to the forward model (i.e., LSTM), and combining both can boost the performance. 
These two LSTM variants are very practical, and they can be deemed as training strategies without alteration of the LSTM fundamentals. 
We also studied some additional LSTM training strategies, which can further improve the accuracy. 
We evaluated our models on two SAR and one non-SAR datasets, and the promising results demonstrated the effectiveness of our approaches in HAR applications.
\end{abstract}


\section{Introduction}
Human activity recognition (HAR) represents one of the core pillars of  mobile, wearable, and pervasive computing, with many practical applications in fields such as health and sleep monitoring \cite{Autism,Liaqat2019realworld, fatigue,Halloran2019,bing2020,Bhattacharya2022Leveraging,Ogbuabor2018healthcare}, behaviour and skill assessment \cite{qingxin2019unsupervised,babystroke,Yoshimura2022work, skill15}, or sports coaching\cite{Cricket,climbAX, phdthesis}, to name but a few.
For traditional machine learning based HAR approaches, feature engineering plays an important role \cite{plotz2011activity,Bulling14,haresamudram2019role,hammerla2013preserving}. 
With expert-designed features, classification can be performed using conventional classifiers such as SVM or KNN \cite{Cricket,Autism}. 
However, designing effective features tends to be a somewhat tedious trial-and-error process, and discriminant features may vary from task to task, making system-developing expensive and less sustainable. 
Recently, deep learning (DL) became the major approach for HAR and behaviour modeling, mainly reasoned by its ability to learn high-level representation in an end-to-end manner \cite{HAR_review18}. 
Based on fixed-length windows/ frames, activity features can be learned via a range of frame-wise DL methods (e.g., RBM \cite{RBM_HAR}, CNN \cite{hammerla2016deep}, DeepConvLSTM \cite{ordonez2016deep}, etc.), with substantial performance improvement when compared with the traditional approaches. 

\begin{table}[t] 
    \centering
    \begin{tabular}{|l|p{6cm}|}
        \hline
        Periodic & Activities or gestures exhibiting periodicity, such as walking,running, rowing, biking, etc.  
        \\ \hline
        Static &Static postures or static pointing gestures. \\ \hline
        Sporadic & The activity or gesture occurs sporadically, interspersed with other activities or gestures.\\ \hline
    \end{tabular}
    \caption{Categorization of activities according to \cite{Bulling14}.}
    \label{tab:activity_type}
    \vspace*{-1em}
\end{table}

In \cite{Bulling14}, activities were categorized into three types: 
\textit{i)} static activities 
\textit{ii)} periodic activities; and 
\textit{iii)} sporadic activities, with related descriptions shown in Table \ref{tab:activity_type}.  
Out of these types, the problem of recognizing static or periodic activities (such as lying, standing, walking, running, cycling, ascending stairs, etc.) has been well-addressed by recent frame-wise DL approaches (e.g., \cite{hammerla2016deep,ordonez2016deep}). 
Recognizing sporadic activities,  
where activities of interest tend to be non-periodic, and occur less frequently in contrast to the large amount of irrelevant background activities,
still remain a major challenge in the HAR research community.
Recent works suggested that sequential DL models (e.g., sample-wise LSTM \cite{hammerla2016deep,guan2017ensembles}) have great potential for modeling the non-periodic activities (e.g., opening the door, closing the drawer, etc.), which is worth exploring on solving the challenging SAR problems.

In this paper, we studied some LSTM training strategies for HAR. 
Specifically, two simple yet effective LSTM variants, namely delay model, and inverse model,  were developed and studied for two SAR scenarios (i.e, with and without time-critical requirement).
For delay model, a time delay interval $\Delta$ is defined, based on which the model  performs LSTM-like, sample-wise classification at timestamp $t$ based on the signal received at and before timestamp $t+\Delta$. 
Performance gains are achieved by taking advantage of the "$\Delta$ timestamps of future information", which can be set by the users for the best trade-off between (decision-making) latency and effectiveness. 

For regular SAR tasks (without time-critical requirement), the second proposed model-- the inverse model--performs sample-wise classification in an inverted processing manner (in contrast to LSTM).
This model learns patterns from the time-series inversely, which, as we found, is complementary to the forward model (i.e., LSTM), and combining them effectively boosts the overall classification accuracy.
Both LSTM variants are simple yet effective, and they can be deemed as training strategies without adding the need to modify the underlying LSTM modeling foundations.
As such, they are universally applicable with virtually no added costs or efforts.

Based on these two LSTM variants, we also studied additional training strategies, i.e., hyper-parameter as variable strategy (HYPAV) and epoch-wise bagging strategy \cite{guan2017ensembles} on some SAR/non-SAR datasets,  and the improved performance suggested they are generic, and can be employed as building blocks for other advanced frameworks in HAR research.

\section{Related Work}
\subsection{Human Activity Recognition(HAR)}
HAR is one of the core research areas in wearable computing, and deep learning (DL) based modeling approaches are now considered the state-of-the-art and the  most effective method for extracting behaviour patterns. 
Classical DL-based HAR works include the application of CNN \cite{yang2015deep}, LSTM\cite{hammerla2016deep}, Bi-directional LSTM(Bi-LSTM) \cite{hammerla2016deep}, Convolutional LSTM\cite{ordonez2016deep}, Attention models~\cite{morales2022acceleration, khaertdinov2021deep,abedin2021attend}, etc. 
To address the problem caused by inconsistent labels, an efficient and flexible modeling approach named dense sampling with full convolution neural network was proposed \cite{DenseLabelling}. 
In \cite{qian2019novel}, a deep learning approach was designed to learn statistical, spatial, and temporal features for HAR.
To improve the performance of HAR, various fusion techniques (e.g., via sensors or classifiers \cite{liu2020giobalfusion,guan2017ensembles}) were also developed. 

Recently, and in addition to, for example, cross-modality transfer techniques \cite{kwon2020imutube,kwon2021approaching,kwon2021complex},  more advanced DL methods were proposed and employed to solve the core challenges, e.g., lack of annotations, large subject variability, scalable training in HAR. 
To address the lack of annotation problem, self-supervised learning were employed in ~\cite{tang2021selfhar, mohamed2022har, self700k_22, harish2022assessing, haresamudram2020masked, haresamudram2021contrastive, haresamudram2022investigating} which can take advantage of the unlabelled data for performance improvement. 
In ~\cite{suh2022adversarial, bai2020adversarial, qian2021latent, Su22_disentanglement}, adversarial learning methods were developed to reduce the variability among subjects for user-independent HAR. 
Federated learning approaches were studied in \cite{presotto2022fedclar,Fed_HAR22} for privacy-preserving training and personalized HAR.

\subsection{Sample-wise Long Short Term Memory (LSTM)}
Most of the DL-base HAR approaches performed frame-wise classification. 
Here we also review sample-wise LSTM, which suggested their effectiveness in modeling sequential activities \cite{hammerla2016deep, guan2017ensembles}. 
Based on LSTM, the proposed LSTM variants can be easily derived.

LSTM is a hidden unit in recurrent neural networks(RNN), and it consists of three gates (i.e., forget gate, input gate, and output gate) and one memory cell \cite{hochreiter1997long}.
For an exemplary one-layer LSTM, it takes current input signal $\boldsymbol{x}_{t}$, the previous hidden and cell state $\boldsymbol{h}_{t-1}$, $\boldsymbol{c}_{t-1}$, and outputs the current hidden and cell state $\boldsymbol{h}_{t}$, $\boldsymbol{c}_{t}$. Specifically, the forward pass can be written as follows:
$$\boldsymbol{f}_t=\sigma \left(\boldsymbol{W}^T_{\boldsymbol{xf}} \boldsymbol{x}_t+\boldsymbol{W}^T_{\boldsymbol{hf}} \boldsymbol{h}_{t-1}+\boldsymbol{b}_{\boldsymbol{f}}\right)$$
$$\boldsymbol{i}_t=\sigma \left(\boldsymbol{W}^T_{\boldsymbol{xi}} \boldsymbol{x}_t+\boldsymbol{W}^T_{\boldsymbol{hi}} \boldsymbol{h}_{t-1}+\boldsymbol{b}_{\boldsymbol{i}}\right)$$
$$\boldsymbol{o}_t=\sigma \left(\boldsymbol{W}^T_{\boldsymbol{xo}} \boldsymbol{x}_t+\boldsymbol{W}^T_{\boldsymbol{ho}} \boldsymbol{h}_{t-1}+\boldsymbol{b}_{\boldsymbol{o}}\right)$$
$$\tilde{\boldsymbol{c}}_t=\tanh \left(\boldsymbol{W}^T_{\boldsymbol{xc}} \boldsymbol{x}_t+\boldsymbol{W}^T_{\boldsymbol{hc}} \boldsymbol{h}_{t-1}+\boldsymbol{b}_{\boldsymbol{c}}\right)$$
$$\boldsymbol{c}_t=\boldsymbol{f}_{t} \odot \boldsymbol{c}_{t-1}+\boldsymbol{i}_t \odot \tilde{\boldsymbol{c}}_t$$
$$\boldsymbol{h}_t=\boldsymbol{o}_t \odot \tanh \left(\boldsymbol{c}_t\right)$$
where $\sigma$ is the sigmoid function;$\odot$ is the entry-wise multiplication;  $\boldsymbol{f}_t, \boldsymbol{i}_t, \boldsymbol{o}_t, \tilde{\boldsymbol{c}}_t$ are the internal outputs of the three gates and the memory cell; $\boldsymbol{c}_t,\boldsymbol{h}_t$ are the outputs of the LSTM unit.
With $\boldsymbol{h}_t$ at timestamp $t$, for $C$ classes (e.g., with label $\boldsymbol{y}_t = \{1, 2,..., C\}$) the output of the sample-wise feed-forward pass can be written as:
\begin{equation} \label{eq:LSTM_prob}
\boldsymbol{p}^{LSTM}(\boldsymbol{y}_t|\boldsymbol{x}_t,\boldsymbol{h}_{t-1},\boldsymbol{c}_{t-1})=\operatorname{softmax}\left(\boldsymbol{W}^T_{\boldsymbol{h} \boldsymbol{C}} \boldsymbol{h}_{t}+\boldsymbol{b}_{\boldsymbol{C}}\right)
\end{equation}
which is the class probability vector. 
Given that,
label $\hat{y}_t$ is assigned to the class with the largest entry, i.e.,
\begin{equation}\label{eq:lstm_baseline}
\hat{y}_t = \arg \max_{\{1,...,C\}} \boldsymbol{p}^{LSTM}(\boldsymbol{y}_t|\boldsymbol{x}_t,\boldsymbol{h}_{t-1},\boldsymbol{c}_{t-1}).
\end{equation}

For model parameters, 
$\boldsymbol{W}_{\boldsymbol{xf}}$, $\boldsymbol{W}_{\boldsymbol{hf}}$, $\boldsymbol{W}_{\boldsymbol{xi}}$, $\boldsymbol{W}_{\boldsymbol{hi}}$, $\boldsymbol{W}_{\boldsymbol{xo}}$, $\boldsymbol{W}_{\boldsymbol{ho}}$,
$\boldsymbol{W}_{\boldsymbol{xc}}$, $\boldsymbol{W}_{\boldsymbol{hc}}$,
$\boldsymbol{W}_{\boldsymbol{hC}}$
are weight matrices, and $\boldsymbol{b}_{\boldsymbol{f}}$,$\boldsymbol{b}_{\boldsymbol{i}}$,$\boldsymbol{b}_{\boldsymbol{o}}$,$\boldsymbol{b}_{\boldsymbol{c}}$, $\boldsymbol{b}_{\boldsymbol{C}}$ are bias vectors, which can be trained using backpropagation through time (BPTT)\cite{hochreiter1997long}.

\subsection{Sample-wise LSTM Ensemble }
In \cite{guan2017ensembles}, epoch-wise bagging scheme was applied to generate a number of base sample-wise LSTM learners to be aggregated for HAR. 
To improve the diversity of ensemble, some hyper-parameters were modelled as variables, e.g., window length, batch size, initial sampling point, which were dynamically changing in an epoch(or batch) manner to generate less correlated epoch-wise base classifiers. 
If we simplify LSTM's class probability vector (i.e., Eq. \ref{eq:LSTM_prob}) as $\boldsymbol{p}^{LSTM}(\boldsymbol{y}_t|\boldsymbol{x}_t)$, for an ensemble of $M$ epoch-wise LSTM classifiers, the aggregated score (via a score-level fusion) can be expressed as:   
\begin{equation}
\boldsymbol{p}^{LSTM}_{fusion}(\boldsymbol{y}_t|\boldsymbol{x}_t) = \frac{1}{M}\Sigma_{m=1}^M \boldsymbol{p}^{LSTM}_m(\boldsymbol{y}_t|\boldsymbol{x}_t),
\end{equation}
where $\boldsymbol{p}^{LSTM}_m(\boldsymbol{y}_t|\boldsymbol{x}_t)$ represents the $m^{th}$ base learner in the LSTM ensemble.
In \cite{guan2017ensembles}, based on the inequality of arithmetic and geometric means, it was proved that if cross entropy (CE) was used, the fusion loss $L^{LSTM}_{fusion}$ is always less or equal than the average loss of base learners $\frac{1}{M}\Sigma_{m=1}^M L^{LSTM}_m$, indicating the benefit of employing ensemble approaches for performance improvement. 
Motivated by this, in this work we also studied the fusion of base learners generated from the proposed LSTM variants.


\begin{figure}[htbp]
\centering
\includegraphics[width=\columnwidth]{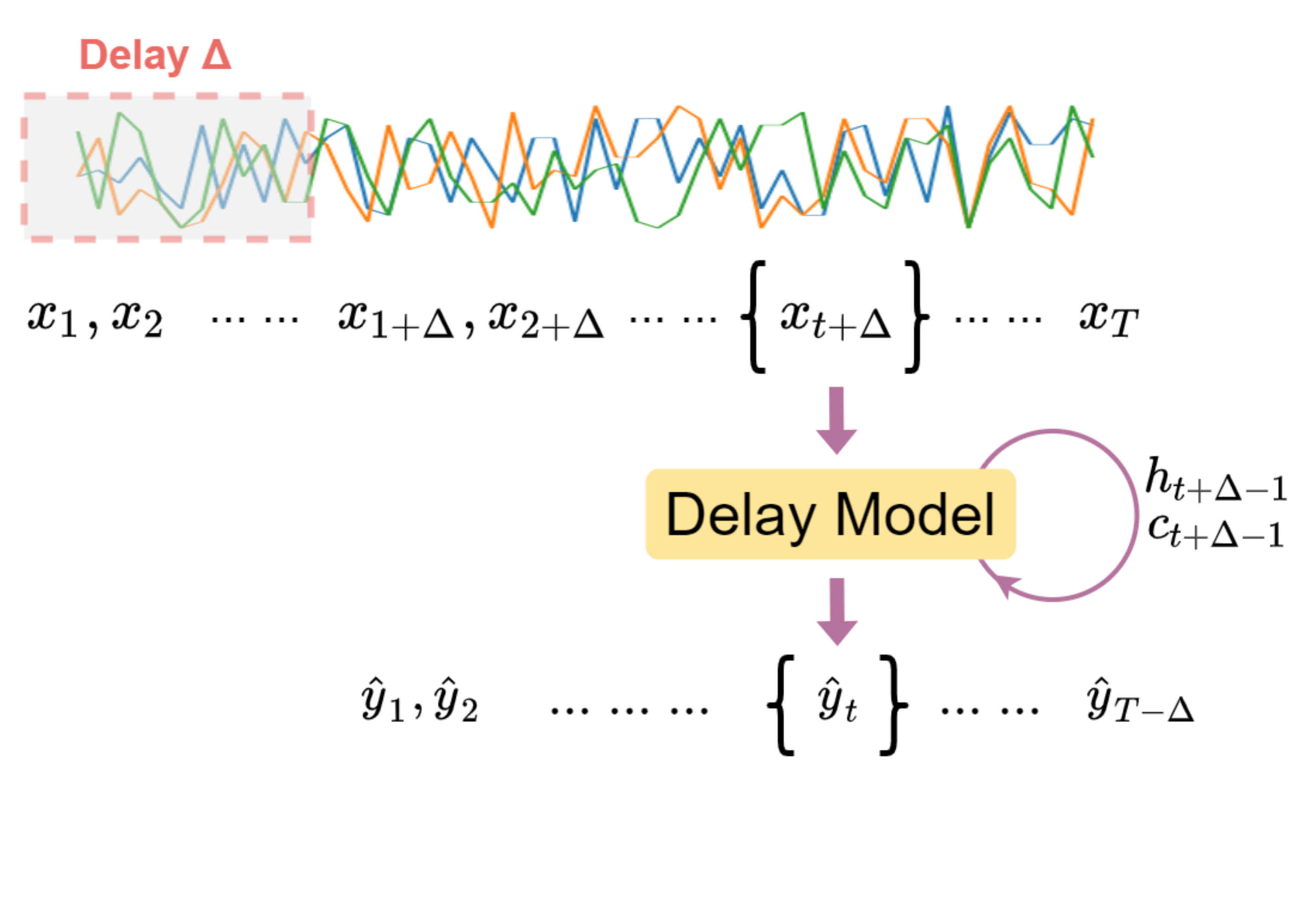}
\caption{The delay LSTM model for time-critical applications (with pre-defined delay interval $\Delta$)}
\label{fig:delay_model}
\end{figure}

\section{Methodology}

In literature, the mainstream DL-base HAR methods performed frame-wise classification. 
Signals within each pre-defined (sliding) window is associated with an activity label, before DL was employed to learn the mapping.
It has achieved its success in the general HAR modeling
since the information of most activities (especially periodic/static types) can be preserved by using short windows (e.g., 1s), and thus can be modelled via the independent window-label pairs. 

However, there might exist prolonged sequential activities to be recognized, which requires more information beyond the length of pre-defined windows.
It was found that sample-wise stateful LSTM (short for LSTM in this paper) can carry information across windows/frames \cite{hammerla2016deep}, with promising results in sequential activities modeling \cite{hammerla2016deep, guan2017ensembles}, making it a potential solution for the challenging SAR tasks.
Moreover, compared with frame-wise approaches, it can perform activity segmentation directly, which might be practical in many applications (e.g., to predict the exact activity duration).

Given that, here we develop some LSTM variants, which can be considered as training strategies without alternation of the LSTM fundamentals.

\subsection{The Delay LSTM Model}
For some time-critical SAR applications such as health monitoring, it is crucial to have the events or incidents detected in low latency. 
For example,  the Freeze-of-Gait incidents from patients with Parkinson's disease should be detected in less than 2s after its onset \cite{Bachlin2010}.
The delay model proposed in this paper can be used to determine the latency for optimal accuracy in such tasks. 

Specifically there is a user-defined hyper-parameter, delay interval ($\Delta$), 
and at timestamp $t+\Delta$ the delay model will output the "delayed predicted label $\hat{y}_t$", as shown in Fig.~\ref{fig:delay_model}.

\begin{figure}[htbp]
\centering
\includegraphics[width=\columnwidth]{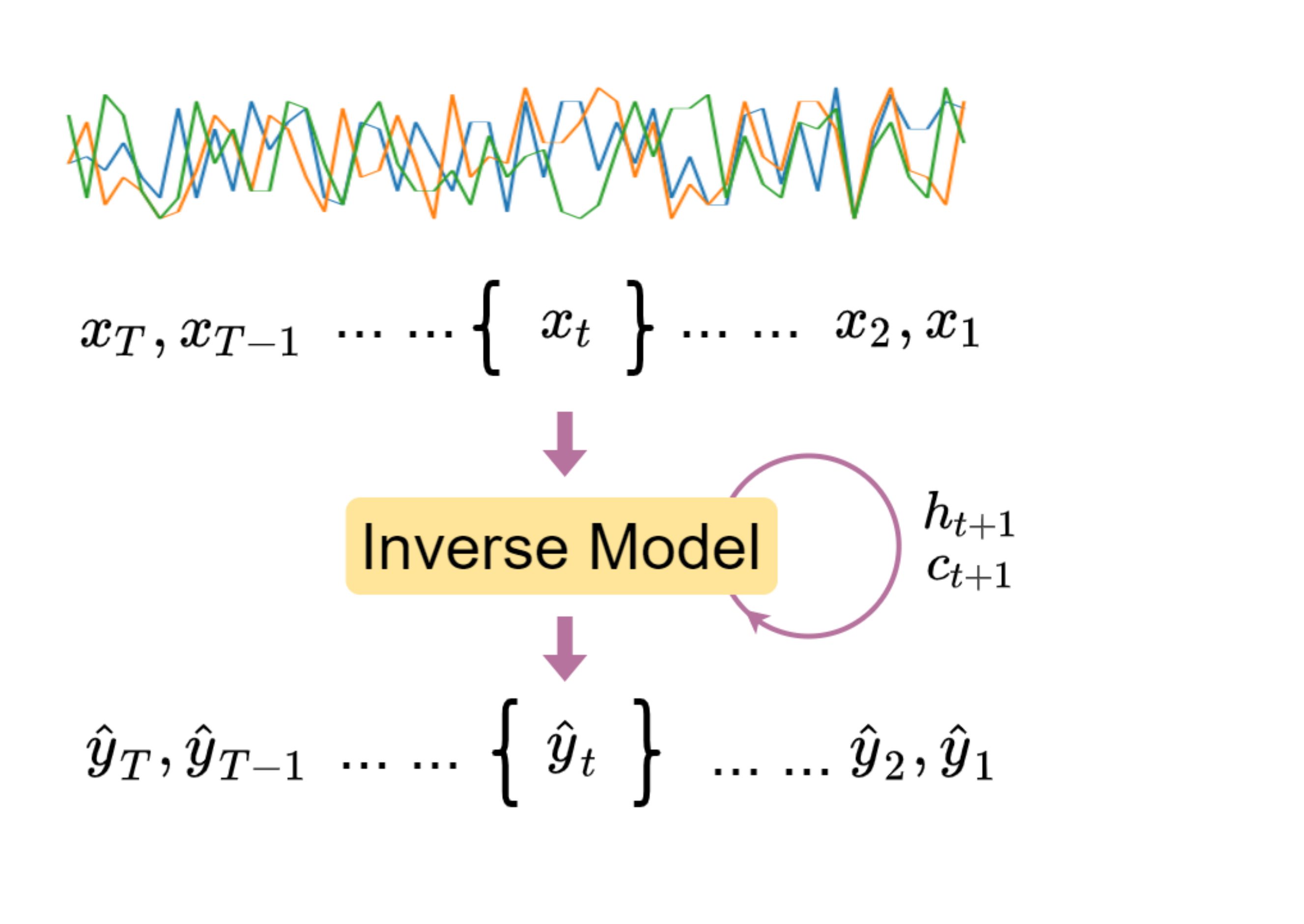}
\caption{The Inverse LSTM Model}
\label{fig:inverse_model}
\end{figure}

In essence, it is to give the model $\Delta$ timestamps of "additional response time" (hence delay model) to better "predict the past".  
Therefore, the delay model is expected to be able to exploit the $\Delta$ timestamps of additional signal information for performance gain.

In contrast to vanilla LSTM, which can map signals to labels at the same timestamp (i.e., from $\boldsymbol{x}_t$ to $\hat{y}_t$ as shown in Eq.\  \ref{eq:lstm_baseline}), the delay LSTM model 
written as
\begin{equation}\label{eq:lstm_delay}
\hat{y}_t = \arg \max_{\{1,...,C\}} \boldsymbol{p}^{DLY}(\boldsymbol{y}_t|\boldsymbol{x}_{t+\Delta},\boldsymbol{h}_{t+\Delta-1},\boldsymbol{c}_{t+\Delta-1})
\end{equation}
maps the current signal $\boldsymbol{x}_{t+\Delta}$ to prediction with $\Delta$ timestamps of delay (i.e.,$\hat{y}_t$). 
In Eq. \ref{eq:lstm_delay}, $\boldsymbol{h}_{(.)}$ and $\boldsymbol{c}_{(.)}$ correspond to hidden and cell states. 
 

\subsection{The Inverse LSTM Model}

Compared with LSTM, which maps signal-label pairs in a forward way, i.e., from the order $\{ (\boldsymbol{x}_1, y_1), ... ,$  
$(\boldsymbol{x}_T, y_T) \}$, the inverse model performs the mapping
in an inverted order, i.e., $\{ (\boldsymbol{x}_T, y_T), ... ,$  
$(\boldsymbol{x}_1, y_1) \}$, aiming at modeling the sequence from a different view.  
Fig. \ref{fig:inverse_model} illustrates the inference process of the inverse model.
For timestamp $t$ we can define it as: 
\begin{equation}\label{eq:lstm_inverse}
\hat{y}_t = \arg \max_{\{1,...,C\}} \boldsymbol{p}^{INV}(\boldsymbol{y}_t|\boldsymbol{x}_{t}, \boldsymbol{h}_{t+1}, \boldsymbol{c}_{t+1}),
\end{equation}
where $\boldsymbol{h}_{t+1}$,$\boldsymbol{c}_{t+1}$ are the "previous" (at timestamp $t+1$) hidden and cell states.
Motivated by the concept of multi-view learning \cite{multiview_survey}, the introduction of inverse model is to analyze the activity signals from a different perspective, i.e., looking at activities from the inverse order, which may be complementary to LSTM. 
For better performance in the challenging SAR tasks, we aggregate the inverse model and LSTM via a score-level fusion:
\begin{equation}\label{eq:lstm_inverse}
\begin{split}
\hat{y}_t = \arg \max_{\{1,...,C\}}  &\frac{1}{2} \{\boldsymbol{p}^{INV}(\boldsymbol{y}_t|\boldsymbol{x}_{t}, \boldsymbol{h}_{t+1}, \boldsymbol{c}_{t+1})\\ &+ \boldsymbol{p}^{LSTM}(\boldsymbol{y}_t|\boldsymbol{x}_{t}, \boldsymbol{h}_{t-1}, \boldsymbol{c}_{t-1})\},
\end{split}
\end{equation}
from which we can see $\boldsymbol{p}^{LSTM}(\boldsymbol{y}_t|\boldsymbol{x}_{t},\boldsymbol{h}_{t-1}, \boldsymbol{c}_{t-1})$ is based on the past hidden/cell states, while $\boldsymbol{p}^{INV}(\boldsymbol{y}_t|\boldsymbol{x}_{t},\boldsymbol{h}_{t+1}, \boldsymbol{c}_{t+1})$ is based on the "future" (i.e., "inverted past") hidden/cell states. 
In this case, it can be beneficial to combine these two complementary information for improved performance.

\begin{figure*}
     \centering
     \begin{subfigure}[b]{0.48\textwidth}
         \centering
         \includegraphics[width=\textwidth]{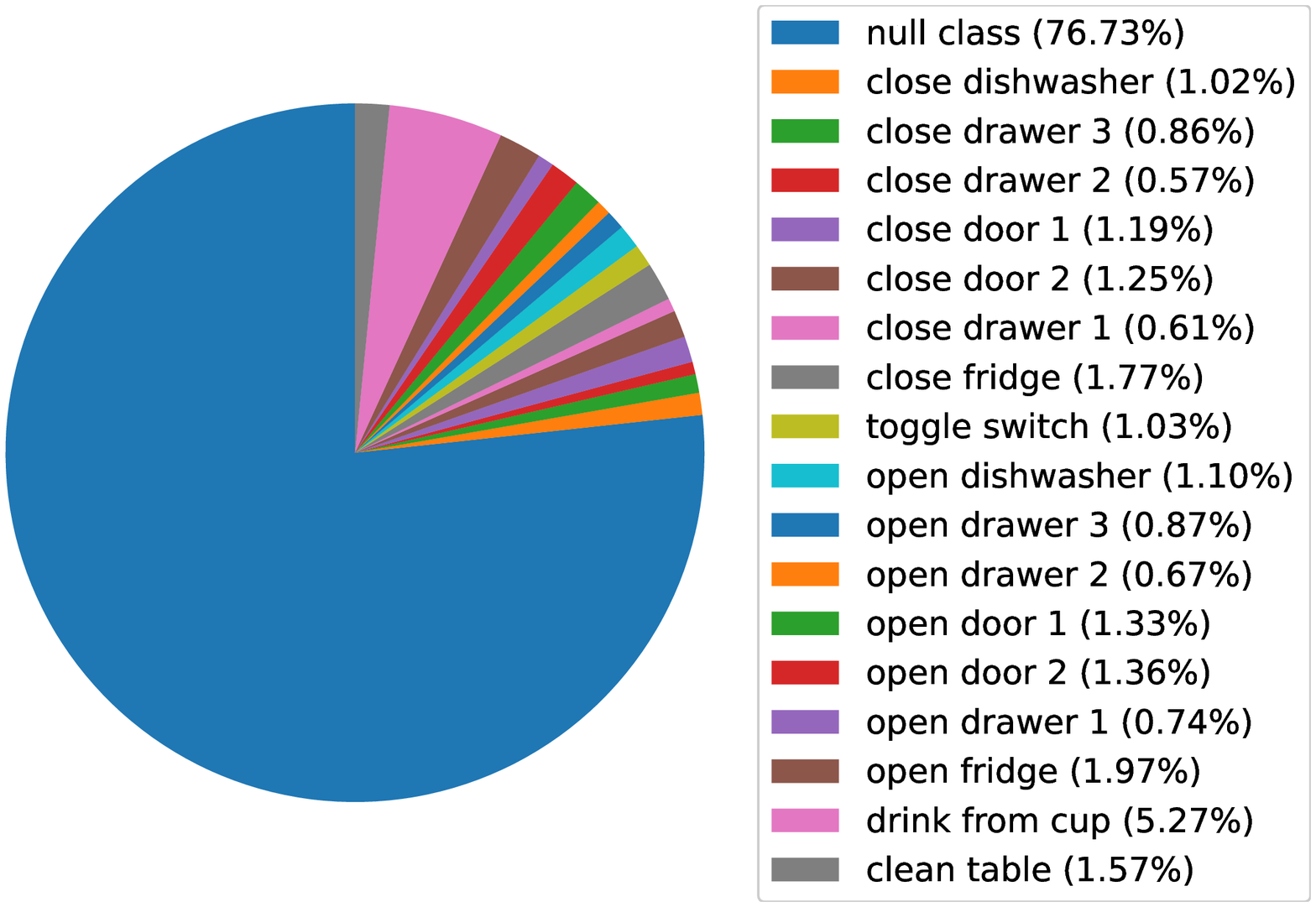}
         \caption{OPP dataset}
         \label{fig:OPP_dataset}
     \end{subfigure}
     \hfill
     \begin{subfigure}[b]{0.45\textwidth}
         \centering
         \includegraphics[width=\textwidth]{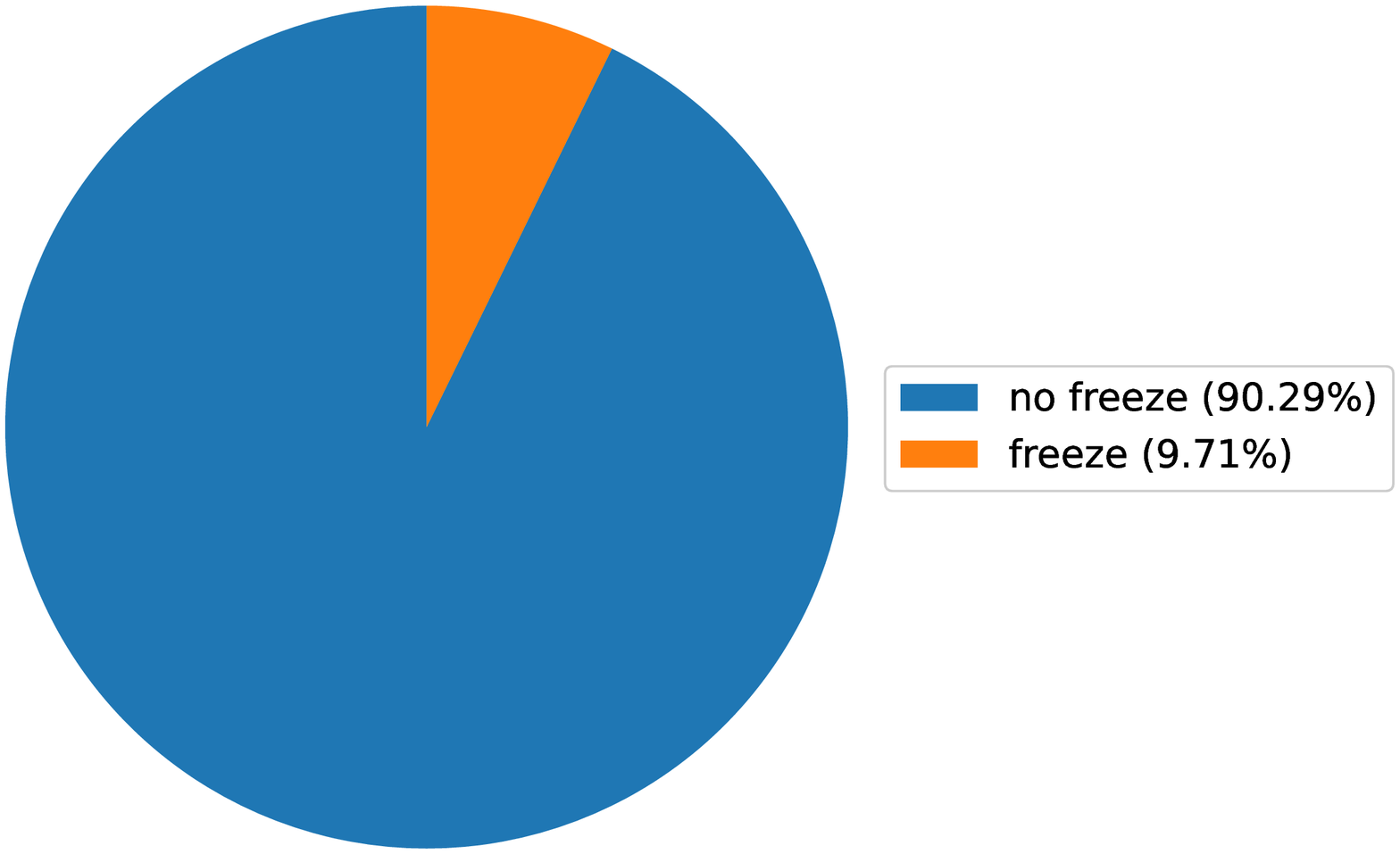}
         \caption{DG dataset}
         \label{fig:DG_dataset}
     \end{subfigure}
     \hfill
     \begin{subfigure}[b]{0.42\textwidth}
         \centering
         \includegraphics[width=\textwidth]{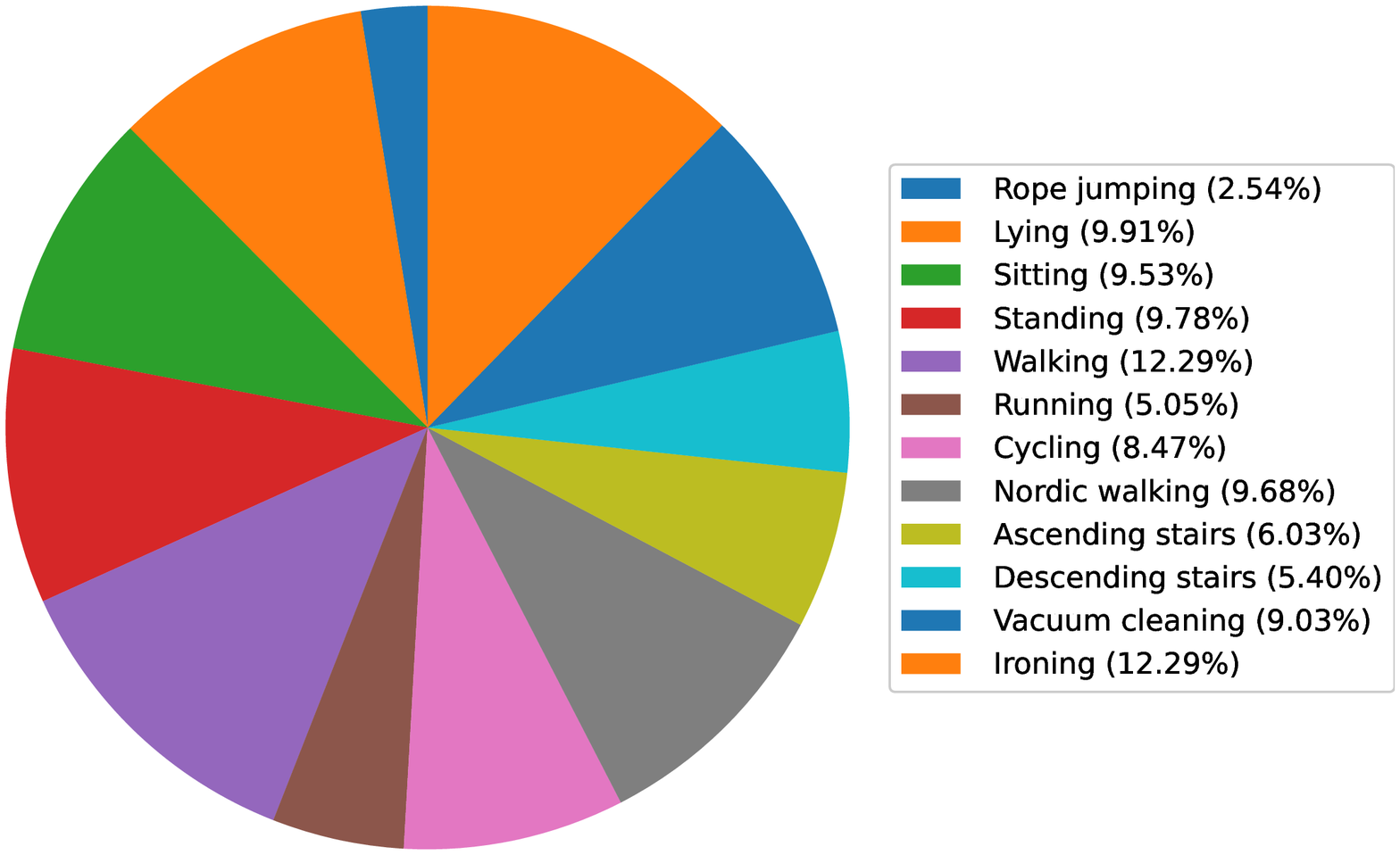}
         \caption{PAMAP2 dataset}
         \label{fig:PAMAP2_dataset}
     \end{subfigure}
        \caption{Class distributions of the three datasets used in this paper; OPP and DG cover SAR tasks, whereas PAMAP2 targets Non-SAR tasks.}
        \label{fig:Dataset_Pie_Chart}
\end{figure*}


\section{Experiments}
To evaluate the effectiveness of the proposed LSTM variants, we conducted experiments on three public HAR datasets, including  
\textit{i)} Opportunity (OPP) \cite{chavarriaga2013opportunity}, 
\textit{ii)} Daphnet Freeze of Gait (DG) \ \cite{Bachlin2010} for SAR tasks, and \textit{iii)} PAMAP2 \cite{Reiss2012IntroducingMonitoring} for Non-SAR tasks.  
Moreover, based on the two LSTM variants, we also studied some additional training strategies, i.e., Hyper-parameter as variable (HYPAV) strategy and Epoch-wise bagging strategy for performance improvement.

\subsection{Datasets}
In our experiments, datasets with both SAR and Non-SAR tasks were used.
Specifically, OPP and DG datasets (corresponding to SAR tasks) cover low-frequent,  sporadic activities, contrasted by large (over-represented) portion of ``other'' (null) class activities.
The resulting class imbalance is substantial: $>75\%$ of the data are NULL for OPP, and $>90\%$ belong to the non-target class in DG, rendering the analysis task  challenging. 
We also evaluated our models on the Non-SAR PAMAP2 dataset -- primarily for robustness and generalization evaluation. 
The class distributions of these three datasets are illustrated in Fig. \ref{fig:Dataset_Pie_Chart}.
These datasets were detailed as follows. 
\subsubsection{Opportunity (OPP)}
OPP \cite{chavarriaga2013opportunity} is probably the most challenging HAR dataset.
From four participants, $17$ pre-defined kitchen activities such as opening the door, closing the drawer, cleaning, etc. as well as the background activity (referred to as the Null class) were recorded.
The pre-defined activities tend to be non-periodic, and most of them cover only  $<2\%$ each of the overall dataset, in contrast to $>75\%$ of the null class.
Following \cite{hammerla2016deep, guan2017ensembles}, we used the 79-dimensional sensor data with sampling rate at 30Hz. 

\begin{figure*}
     \centering
     \begin{subfigure}[b]{0.45\textwidth}
         \centering
         \includegraphics[width=\textwidth]{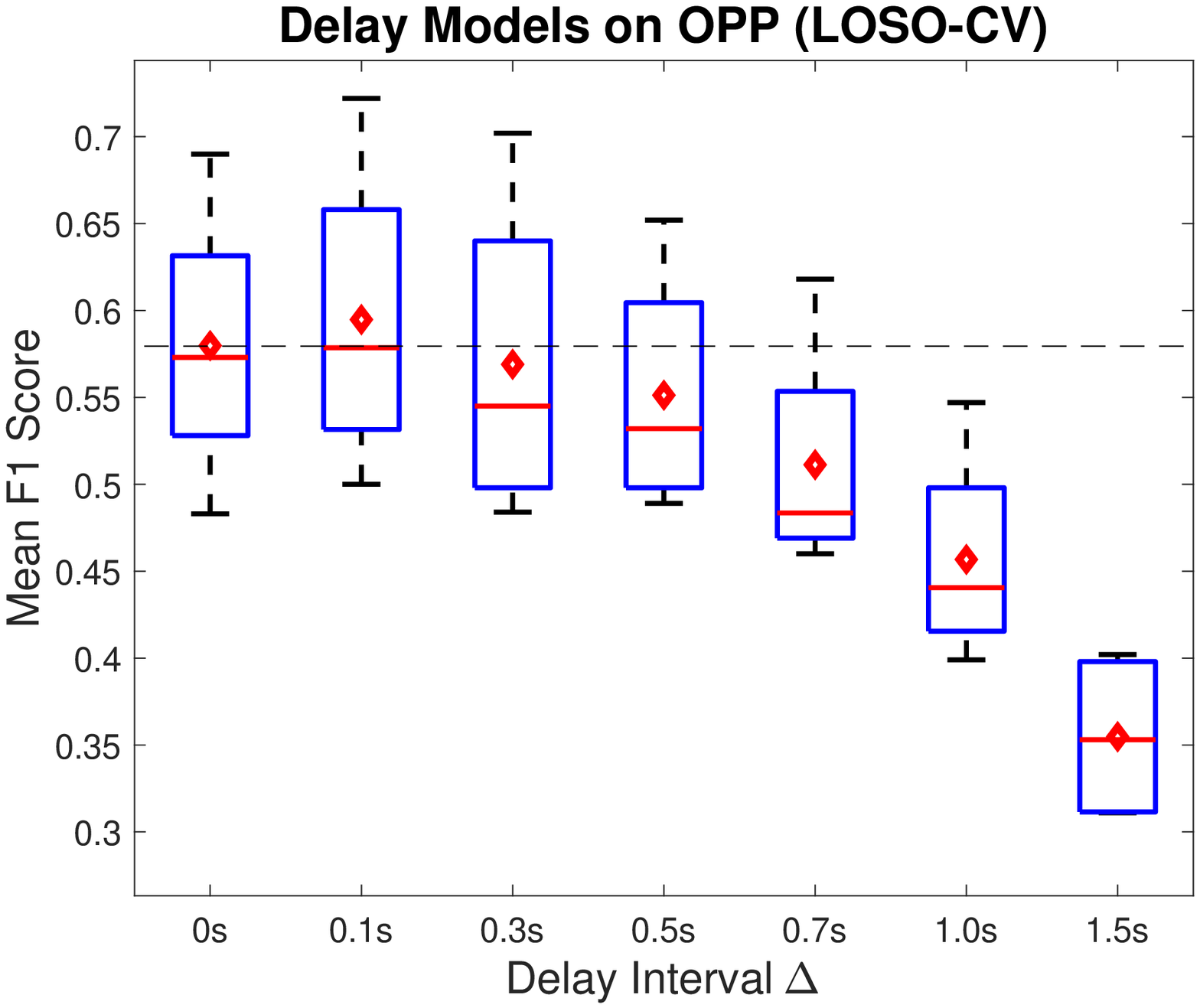}
         \caption{on OPP dataset (regular SAR)}
         \label{fig:Delay_LOSO_OPP}
     \end{subfigure}
     \hfill
     \begin{subfigure}[b]{0.45\textwidth}
         \centering
         \includegraphics[width=\textwidth]{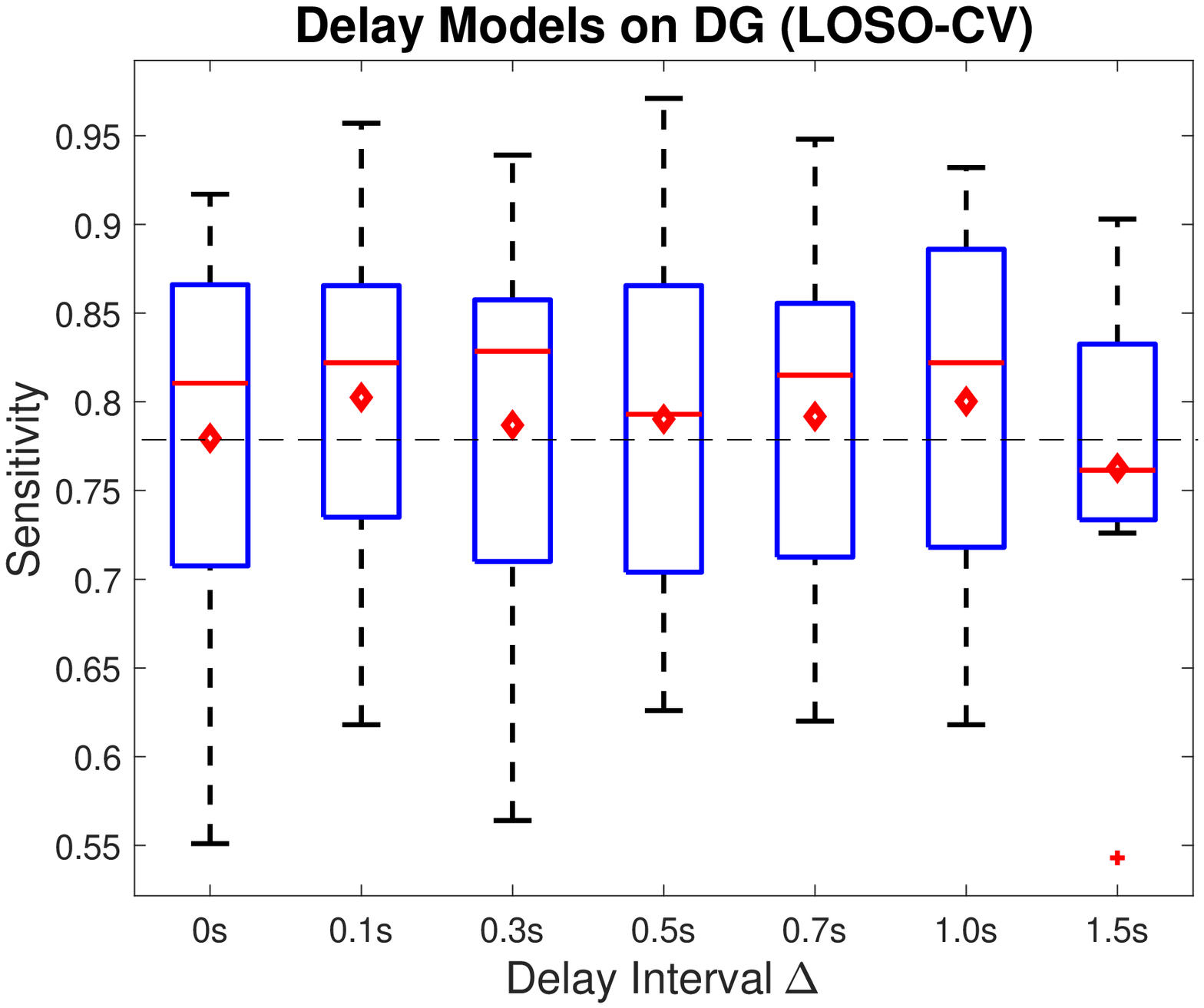}
         \caption{on DG dataset (time-critical SAR)}
         \label{fig:Delay_LOSO_DG}
     \end{subfigure}
     \hfill
     \begin{subfigure}[b]{0.45\textwidth}
         \centering
         \includegraphics[width=\textwidth]{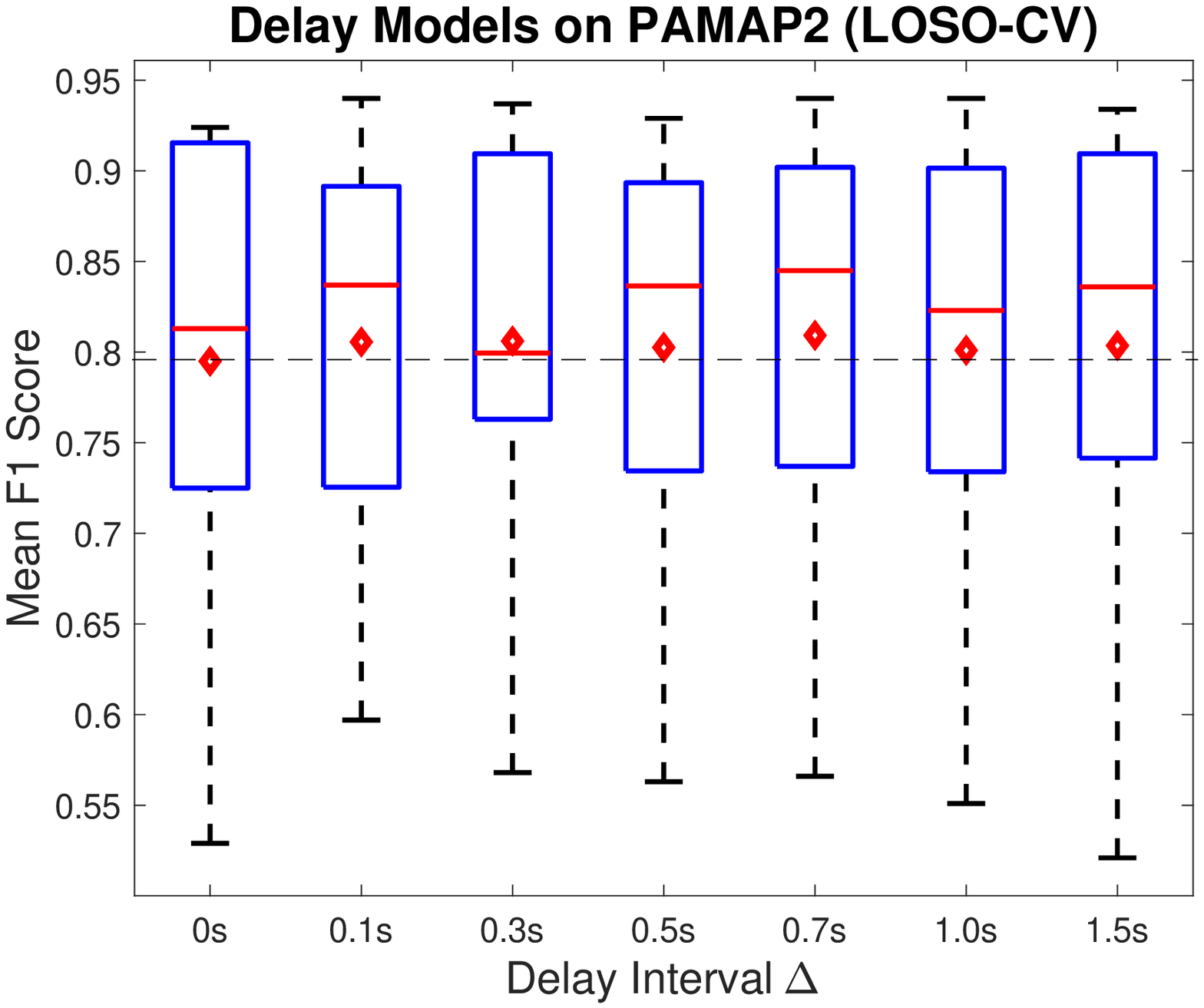}
         \caption{on PAMAP2 dataset (Non-SAR)}
         \label{fig:Delay_LOSO_PAMAP2}
     \end{subfigure}
        \caption{Box-plots of delay models' performance on three datasets in leave-one-subject-out cross validation (LOSO-CV) setting; Mean-F1 scores were used on OPP and PAMAP2, while sensitivity (with specificity fixed at $0.9$) was used for DG dataset; 
        The red circles indicate the average/mean values, and the horizontal dashed line is the average performance of LSTM model (i.e., with delay interval $\Delta=0$) }
        \label{fig:Delay_LOSO}
\end{figure*}

\subsubsection{Daphnet Freeze of Gait (DG)}
In contrast to OPP dataset which contains regular daily sporadic activities, DG is a health monitoring dataset including time-critical SAR tasks. 
The sporadic activity in DG is Freeze-of-Gait (FoG) symptom \cite{Bachlin2010}, which is a common complication in Parkinson's Disease(PD), with sudden and transient inability to walk.
In order to recognize the FoG incidents from the normal walking or other activities (i.e., no freeze class), ten PD patients who suffered from FoG were recruited, and wearable sensors were placed above the ankle, knee, and waist to collect the 9-dimensional accelerometer data \cite{Bachlin2010}.

\subsubsection{PAMAP2}
For Non-SAR tasks, we used the popular PAMAP2 dataset \cite{Reiss2012IntroducingMonitoring}, which includes 12 periodic/static activities such as walking, running, cycling, lying, sitting, etc.
From nine subjects, IMU(including accelerometer, gyroscope, magnetometer), temperature, and heart rate data were recorded via body-worn sensors attached to hand, chest and ankle, yielding a 52-dimensional time-series dataset. 
We downsampled the dataset to 33Hz in our experiments.

\subsection{Settings}

\subsubsection{Evaluation Protocols}
In this work, we evaluated the models primarily using hold-out validation.  
In addition, leave one subject out cross validation (LOSO-CV) was also applied in some experiments as supplementary .

For OPP dataset, in the LOSO-CV setting data from all four participants were used; while in the hold-out setting, we used the protocol defined in \cite{hammerla2016deep, guan2017ensembles}. 
That is, the second run from subject 1 was used as validation data, runs 4 and 5 from subjects 2 and 3 were used as test set, and the remaining data were used for training.

For DG dataset, since subject 4 and 10 do not have the FoG incidents, they were excluded in our experiments.
Therefore, data from the rest eight subjects were used in both LOSO-CV and hold-out settings.
In the hold-out setting, we followed the protocol used in \cite{hammerla2016deep}. 
That is, we used run 1 from subject 9 as validation set, runs 1 and 2 from subject 2 as test set, and used the rest for training.

For PAMAP2 dataset, 
in LOSO-CV, eight subjects were used, and in the hold-out setting we followed the protocol used in  \cite{hammerla2016deep}\cite{guan2017ensembles}. 
That is, we used runs 1 and 2 from subject 5 for validation, and runs 1 and 2 from subject 6 as test while using the rest data for training. 

\begin{table*}[htbp]
  \centering
    \begin{tabular}{|c|c|c|c|c|}
    \hline
     \multicolumn{1}{|l|}{\hspace{25pt}-}&
     \multicolumn{1}{|c|}{OPP \hspace{1pt}}&
     \multicolumn{2}{|c|}{DG \hspace{2pt} }&
     \multicolumn{1}{|c|}{PAMAP2}\\
    \hline 
    Metric  & mean-F1 & Sensitivity & mean-F1  & mean-F1 \bigstrut\\
    \hline    
    LSTM  & 0.673 & 0.821 &  0.668  & 0.795 \bigstrut\\ \hline  
    Delay($\Delta=0.1s$)& 0.669  & 0.832 & 0.671  & 0.798 \bigstrut\\ \hline  
    Delay($\Delta=0.3s$)& 0.678  & 0.834 & 0.668  & 0.801 \bigstrut\\ \hline  
    Delay($\Delta=0.5s$)& 0.688  & 0.848  & 0.683  & 0.804 \bigstrut\\ \hline  
    Delay($\Delta=0.7s$)& 0.685  & 0.839 & 0.684 & 0.806 \bigstrut\\ \hline  
    Delay($\Delta=1.0s$)&  0.667  &0.812  & 0.662  & 0.802 \bigstrut\\ \hline  
    Delay($\Delta=1.5s$)&  0.645 &0.776  & 0.642 & 0.796 \bigstrut\\ 
    \hline
    \end{tabular}%
    \caption{Performance of Delay models in hold-out setting on three datasets}
    \label{tab:Delay_holdout}%
\end{table*}%

To measure the performance, for OPP and PAMAP2 datasets we used the popular mean-F1 score. 
For DG dataset, since the FoG detection task is a typical binary health monitoring problem, which requires suitable thresholds to minimize the false negative predictions, we used sensitivity (with a fixed specificity $0.9$) as the main metric, and we also reported the corresponding mean-F1 score to make it in line with other HAR literature.  

Since we focus on sample-wise activity recognition (as per \cite{hammerla2016deep,guan2017ensembles}), for any frame-wise baselines (e.g., CNN \cite{yang2015deep}, DeepConvLSTM \cite{ordonez2016deep}), we applied a post-processing procedure that maps the frame-wise prediction to sample-level results.

\subsubsection{Network Architectures and Hyper-parameters}
For comparison to the unmodified LSTM baseline, we followed \cite{guan2017ensembles} and used a two-layer LSTM, with 256 hidden units for each layer.
Dropout was used for both the first/ second hidden layers with a probability of $50\%$. 
Cross entropy loss was used and the learning rate was set to $0.001$.
In the default settings (i.e., when without HYPAV strategy), we set the batch size to $128$, and window length to $1$ second for all datasets.

\subsection{Delay Models}

Based on LOSO-CV setting, in Fig. \ref{fig:Delay_LOSO} we reported the delay models' performance using box-plots. 
Specifically, on the three datasets (OPP, DG, and PAMAP2) we showed the results corresponding to different delay intervals $\Delta=\{0s, 0.1s, 0.3s, 0.5s, 0.7s, 1s, 1.5s\}$.
Note LSTM model is a special case when we set $\Delta=0$ in the delay model. 
In Fig. \ref{fig:Delay_LOSO}, we also reported the average performance using red circles, and plotted LSTM's average performance in dashed line to compare with other delay models.   

For SAR tasks, in Fig. \ref{fig:Delay_LOSO_OPP}, and \ref{fig:Delay_LOSO_DG}, we can observe when there is a small delay interval $\Delta=0.1s$, the delay mechanism can generally improve the average performance, indicating the "short-term future information"  is important to the sequential modeling in SAR tasks. 
However, the results may start to drop when a larger $\Delta$ was applied, suggesting information from the "long future" may be less relevant.  
This trend is more obvious on OPP dataset than DG dataset. 

On the other hand, on the Non-SAR PAMAP2 dataset, performance does not vary much when with different $\Delta$ (as shown in Fig. \ref{fig:Delay_LOSO_PAMAP2}), indicating the contextual information may not be as useful as the ones in SAR tasks.
Nevertheless, they are not harmful to the performance, indicating the generalization capability of the delay models.

We also ran these delay models in hold-out settings on these three datasets, and the results were reported in Table \ref{tab:Delay_holdout}. 
We see the performance is generally consistent with the LOSO-CV settings. 
That is, the performance gains can be achieved when suitable delay intervals were used in SAR tasks (i.e., on OPP and DG datasets), while this trend is not obvious for the Non-SAR PAMAP2 dataset.
These results suggest that the periodic or static activities (i.e., in Non-SAR tasks) can't provide as much sequential information as the sporadic activities for delay models. 

In SAR tasks, it is interesting to see that the optimal delay interval $\Delta$ may vary.
For example, in LOSO-CV settings, best (average) performance can be achieved when $\Delta=0.1s$ on OPP and DG datasets, while in hold-out setting the optimal results were based on $\Delta=0.5s$ or $\Delta=0.7s$.
One possible explanation can be $\Delta=0.1s$ is the optimal for the whole population (i.e., the subjects in the dataset), which is not necessarily the best one for the unseen individuals. 
This indicates that personalized or adaptive delay interval should employed, which will be studied in the future.

In practical time-critical health monitoring scenarios (e.g., on DG dataset), we can see the benefits from the delay models.
In both LOSO-CV and hold-out settings, the detection rate (i.e., sensitivity) can be improved when with small delay intervals (e.g., $\Delta = \{0.1s, 0.3s, 0.5s, 0.7s\}$), indicating delay models can take advantage of the additional contextual information at a cost of short latency $\Delta$.
Since the FoG incidents should be detected as early as possible (with an acceptable tolerance $< 2s$ \cite{Bachlin2010}), our delay model (i.e., with $\Delta \le 0.7s$)) can be served as an effective health monitoring system in this time-critical SAR application. 

For the rest of the paper, for simplicity all experimental results reported were based on hold-out settings.

\subsection{Combining LSTM and Inverse Models}

We also ran the inverse LSTM model on three datasets, which can perform sequential labelling in the backward way. 
Compared with the forward sequential labelling models such as LSTM, inverse model look at the same data from a different view, which may be complementary.
Hence, we combined both models using a simple score-level fusion, and compared it with baseline LSTM and delay models (with the optimal $\Delta$).

\begin{table}[htbp]
  \centering
    \begin{tabular}{|c|c|c|c|c|}
    \hline
     \multicolumn{1}{|l|}{\hspace{25pt}-}&
     \multicolumn{1}{|c|}{OPP \hspace{1pt}}&
     \multicolumn{2}{|c|}{DG \hspace{2pt} }&
     \multicolumn{1}{|c|}{PAMAP2}\\
    \hline 
    Metric  & mean-F1 & Sensitivity & mean-F1  & mean-F1 \bigstrut\\
    \hline    
    LSTM  & 0.673 & 0.821 &  0.668  & 0.795 \bigstrut\\ \hline 
    Delay  & 0.688 & \bf{0.848} &  \bf{0.683}  & \bf{0.806} \bigstrut\\ \hline 
    LSTM$\&$Inverse& \bf{0.705}  & 0.832 & 0.674  & 0.804 \bigstrut\\ \hline  

    \end{tabular}%
    \caption{Performance of LSTM$\&$Inverse models in hold-out setting on three datasets; For delay models, only the ones with the best delay intervals were reported here.}
    \label{tab:INV_holdout}%
\end{table}%

From Table \ref{tab:INV_holdout} we can see on OPP dataset, the mean F1 score can be improved from $0.673$ (LSTM) to $0.705$ (LSTM$\&$Inverse), and the combined model is also $1.7\%$ higher than the best delay model.
On DG and PAMAP2 datasets, however, the performance gains are not as significant as the simple delay models. 
Moreover, compared with LSTM or delay models, the combined LSTM$\&$Inverse sacrifices the (near-)real-time property since the whole inference sequence should be acquired before inverse model can be applied. 
Given that, it is less practical for time-critical health monitoring tasks.

Nevertheless, from the modeling perspective we can see the benefit of employing inverse model under the fusion scheme, which motivates us to explore more fusion strategies on these LSTM variants.

\subsection{Additional Training Strategies}
The proposed LSTM variants can be deemed as training strategies without alteration of the LSTM fundamentals, making them very practical. 
Moreover, we studied additional training strategies for performance improvement, i.e., \textit{i)} Hyper-parameter as variable (HYPAV) strategy and \textit{ii)} Epoch-wise bagging strategy.

\subsubsection{Hyper-parameter as variable (HYPAV) strategy}
In previous sample-wise HAR work \cite{guan2017ensembles}, some hyper-parameters such as window length, batch size were modelled as random variables, which were dynamically changing in different training batches or epochs in order to increase the diversity of the LSTM ensemble system for higher performance.

Different from \cite{guan2017ensembles} which has an ensemble diversity emphasis, here we mainly study how the proposed single LSTM variants can benefit from the "hyper-parameter as variable (HYPAV)" strategy.

Following \cite{guan2017ensembles}, we set window length, batch size, and the sampling position as random numbers. 
Due to these introduced random factors, we ran each model 30 times with the corresponding results reported in the form of mean$\pm$std or box-plotting.
In some experiments, we further employed two-tailed independent t-tests and reported significance levels through p-values, where p $\leq$ 0.05, p $\leq$ 0.01 and p $\leq$ 0.001 correspond to *, **, and ***, respectively.

\begin{figure}[htbp]
\centering
\includegraphics[width=9cm,height=6.5cm]{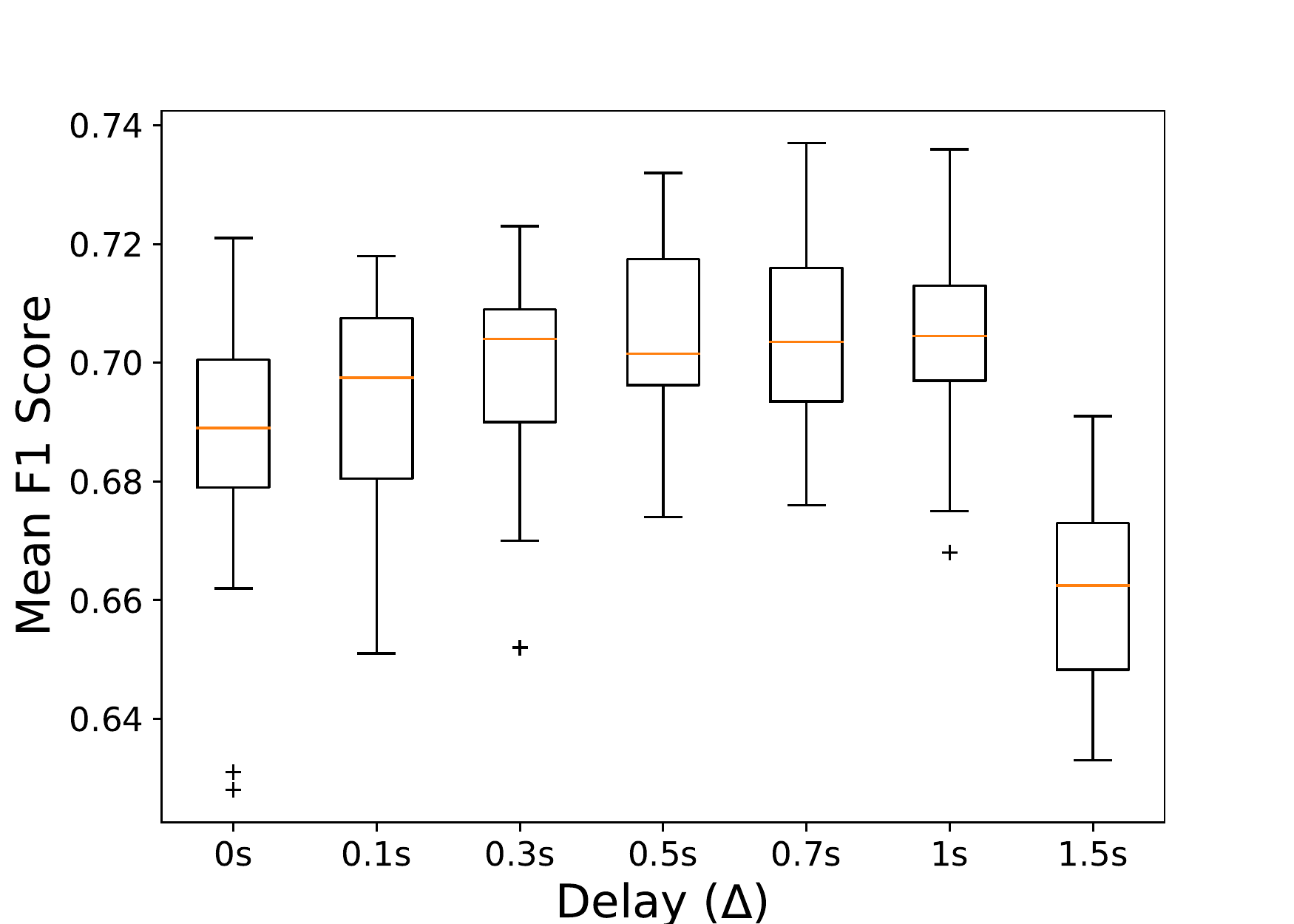}
\caption{Box-plot of mean-F1 w.r.t. $\Delta$ of the delay models on OPP.}
\label{fig:delay_opp}
\end{figure}

Fig.~\ref{fig:delay_opp} shows the box-plot of performance w.r.t. delay interval ($\Delta$) of our delay model on OPP dataset when HYPAV strategy was used, and we can see the performance gain when with $\Delta\le 1s$. 
Similar to the previous results where HYPAV strategy was not used, 
it shows that a longer delay (e.g., $1.5s$) may not contribute positively to the performance. 
This is sensible since signal and label may be less correlated for timestamps far-apart. 
Specifically, when with a short delay $\Delta$, the delay model can effectively exploit the additional contextual information for improved results, and with $\Delta=0.7s$ it has the best mean-F1 $0.705\pm 0.02$, which is nearly $2\%$ higher than the best one with fixed hyper-parameters (i.e., as shown in Table \ref{tab:Delay_holdout}), suggesting the effectiveness of the HYPAV strategy in delay models.

\begin{figure}[htbp]
\centering
\includegraphics[width=6cm,height=5cm]{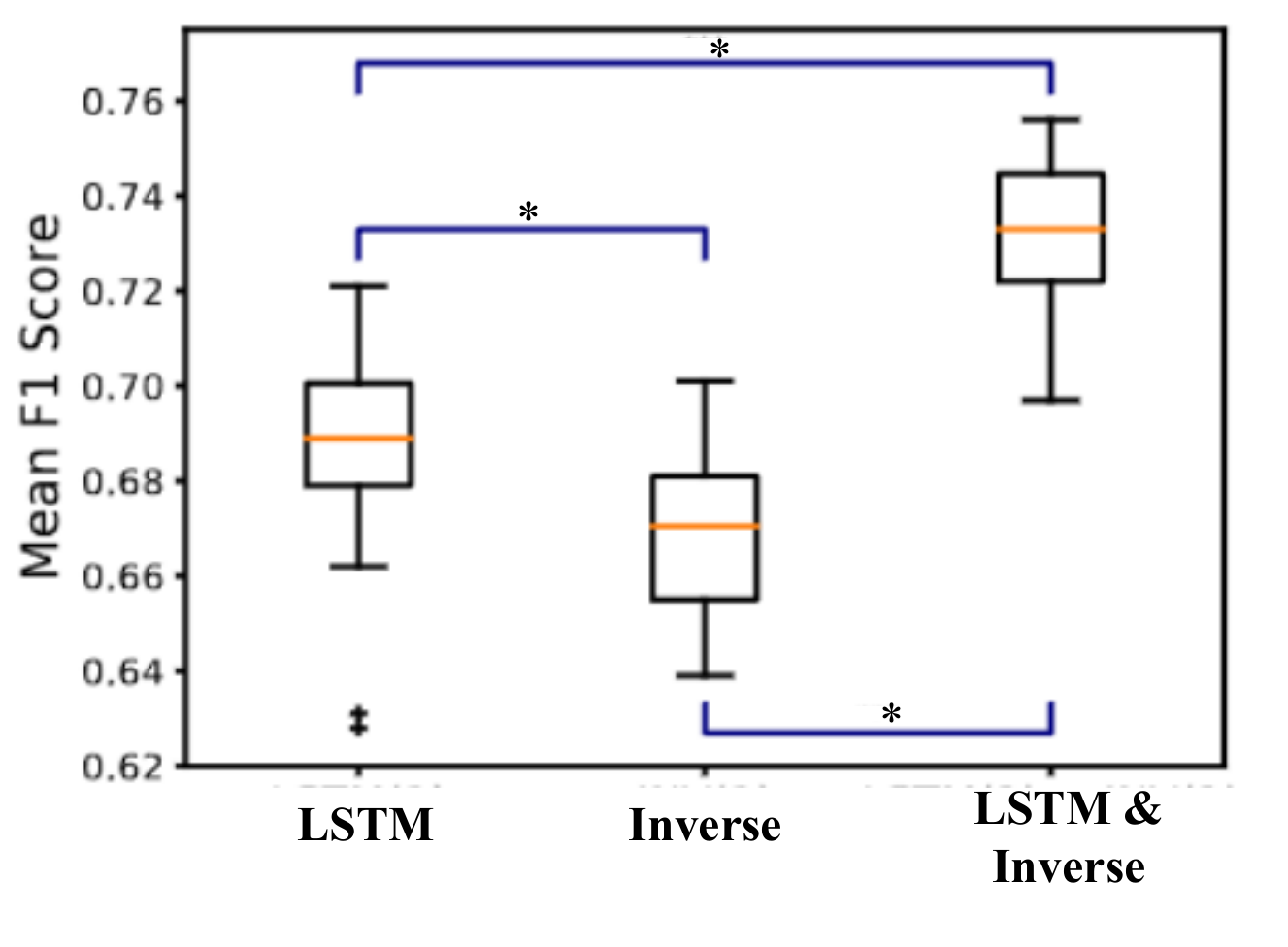}
\caption{Box-plot of mean-F1 scores for LSTM, the inverse model, and their fusion (LSTM$\&$Inverse) on OPP dataset; $*$ indicates p-value $p<0.05$.}
\label{fig:inverse_single_boxplot}
\end{figure}

We also applied HYPAV training strategy to inverse model on OPP dataset, and report the related results in Fig.~\ref{fig:inverse_single_boxplot}. 
Although we do not observe any performance gain (over LSTM) if only inverse model was applied, under HYPAV strategy it shows its powerful complementary capability with LSTM. 
By performing a simple score-level fusion, we can see the combined model LSTM$\&$Inverse achieves mean-F1 $0.733\pm 0.01$, much higher than single models (with p-value $p<0.05$).
Compared with the one with fixed window length and batch size (with mean F1 score $0.705$ as reported in Table \ref{tab:INV_holdout}), this HYPAV training strategy can also bring nearly $3\%$ performance gain in the LSTM$\&$Inverse scheme.

Similarly, we also conducted experiments on DG and PAMAP2 datasets, and the results of these three datasets were summarized in Table \ref{tab:HYPAV}. 
For delay models with multiple results, only the best ones were reported, and we can see the our models can benefit from the HYPAV training strategies, and performance gain can be achieved if the hyper-parameters (e.g., window length, batch size) were modelled as variables. 

\begin{table*}[htbp]
  \centering
    \begin{tabular}{|c|c|c|c|c|c|c|}
    \hline
     \multicolumn{1}{|l|}{\hspace{25pt}-}&
     \multicolumn{2}{|c|}{OPP \hspace{2pt}(mean-F1)}&
     \multicolumn{2}{|c|}{DG \hspace{2pt} (Sensitivity / mean-F1)}&
     \multicolumn{2}{|c|}{PAMAP2\hspace{2pt}(mean-F1)}\\
     \hline 
    Strategy & w/ HYPAV & w/o   & w/ HYPAV & w/o  & w/ HYPAV & w/o \bigstrut\\
    \hline \hline
    Delay Model  & \bf{0.705}$\pm$0.02 & 0.688  & \bf{0.865}$\pm$0.04 \hspace{1pt}/ \bf{0.719}$\pm$0.01 & 0.848 \hspace{1pt}/0.683  & \bf{0.829}$\pm$0.04 & 0.806 \bigstrut\\
    \hline 
    LSTM$\&$Inverse  & \bf{0.733}$\pm$0.01 & 0.705  & \bf{0.833}$\pm$0.03 \hspace{1pt}/ \bf{0.702}$\pm$0.01  & 0.832 \hspace{1pt}/0.674  & \bf{0.829}$\pm$0.05 & 0.804 \bigstrut\\    
    \hline
    \end{tabular}%
    \caption{Performance evaluation of the Hyper-parameter as variable (HYPAV) strategy; For DG dataset, both sensitivity (with specificity fixed at 0.9) and the corresponding mean F1 scores were reported; For delay models with different delay intervals $\Delta$, only the best results were presented here}
    \label{tab:HYPAV}%
\end{table*}%

\begin{figure*}[htbp]
\centering
\includegraphics[width=11cm,height=8cm]{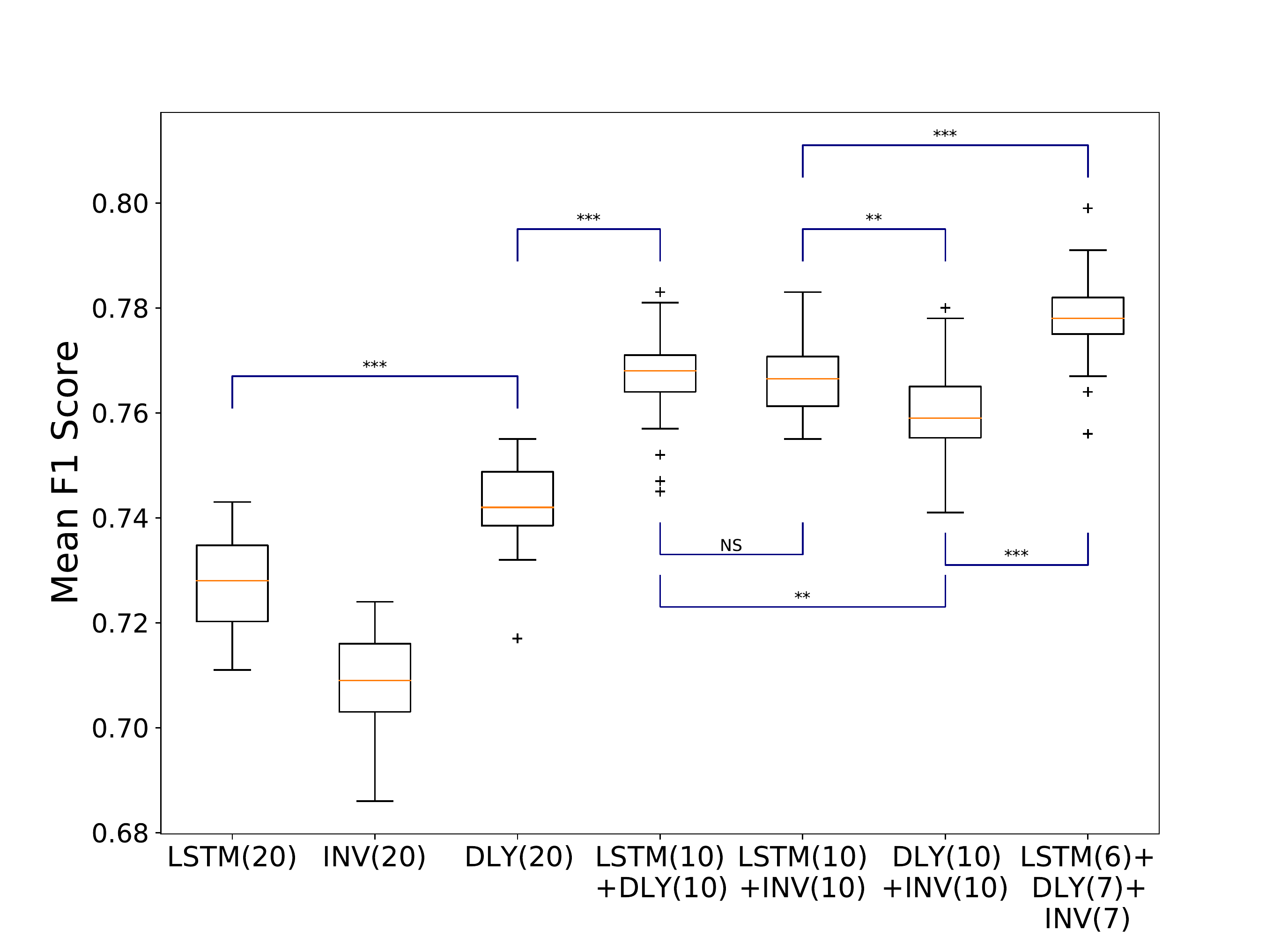}
\caption{Box-plot of mean-F1 scores for different ensembles on Opp; DLY/INV represents the delay/inverse models.}
\label{fig:comparison_fusion_boxplot}
\end{figure*}

\subsubsection{Epoch-wise bagging strategy}

The delay model is very practical, and we can easily achieve some performance gain at a cost of (the pre-defined) latency, making it is a flexible tool for near-real-time prediction applications.
On the other hand, the results also suggested the effectiveness of using score-level fusion of two different LSTMs (e.g., LSTM$\&$Inverse), which motivates us to employ ensemble or more fusion approaches for additional performance gain. 
In \cite{guan2017ensembles}, Guan and Ploetz proposed an epoch-wise bagging scheme for LSTM ensemble, and with the HYPAV strategy, the models trained from different epochs were used as base learners to be aggregated.

Here on OPP dataset, we also studied this epoch-wise bagging framework (with HYPAV) based on the three LSTM types (i.e., LSTM/Delay(DLY)/Inverse(INV)). 
Specifically, we generated different ensembles (with a fixed size of 20 base classifiers) and compared their results in Fig. \ref{fig:comparison_fusion_boxplot}, from which we can observe:
\begin{itemize}
\item for single source ensemble, DLY(20) has much better results than other LSTM(20) or INV(20)
\item multi-source ensembles tend to have higher performance, and the best results can be achieved when fusing the base classifiers from three sources LSTM(6)+DLY(7)+INV(7), significantly higher than any single/dual-source ensembles. 
\end{itemize}

Although base learner’s performance is important, it is also crucial to combine diversified base classifiers. 
Although HYPAV strategy can be used to inject diversity in the epoch-wise bagging scheme, the base learners are derived from the same training process and still are highly correlated \cite{guan2017ensembles}.  
On the other hand, the multi-source ensemble in Fig. \ref{fig:comparison_fusion_boxplot} are derived from three different training processes (i.e., based on LSTM/Inverse/Delay models) which may yield diversified base learners with improved results. 
These results indicate that our proposed LSTM variants can be used as effective building blocks of other advanced schemes for performance boosting.

\subsection{Model Comparison}
On the three HAR datasets, we also compare our LSTM models (Delay and LSTM$\&$Inverse, and multi-source ensemble) with other state-of-the-art approaches.

In Table~\ref{tab:opp_result} we listed the results of other approaches on OPP dataset, including
Dense Labeling \cite{DenseLabelling},  
DeepConvLSTM \cite{ordonez2016deep}, LSTM-S \cite{hammerla2016deep}, Bi-LSTM \cite{hammerla2016deep}, Attention Model \cite{attention18}, LSTM Ensemble \cite{guan2017ensembles}, and Attend$\&$Discriminate\cite{Attend_Discriminate}.
We can see our LSTM variants (Delay ($\Delta=0.7s$) and LSTM$\&$Inverse) yield very promising results, indicating their effectiveness.
They are very practical due to the simplicity nature, and with the epoch-wise bagging strategy (including HYPAV), these LSTM variants can be combined with the highest result on the OPP dataset, which was considered as one of the most challenging dataset. 

We also compared our approaches with other baselines on DG and PAMAP2 datasets, and we can see substantial performance gain when with our LSTM variants and training strategies.  
It is worth noting that on DG dataset, although single delay model had lower performance than the multi-source ensemble, it can be performed in near-real-time (with a delay interval of $0.5s$), making it a practical solution in health monitoring applications. 
On the other hand, on the Non-SAR PAMAP2 dataset, although previous results suggested no significant performance gains can be achieved by using LSTM variants only (as shown in Table \ref{tab:INV_holdout}), we can see the benefits of applying the additional training scheme (i.e., epoch-wise bagging with HYPAV) to these LSTM variants, suggesting their generalization capability and robustness.

\begin{table}[htbp]
\centering
\begin{tabular}{||l|c||}
\hline
Model  & Mean F1 \\ \hline
Dense Labelling \cite{DenseLabelling} &0.624 \\ \hline
DeepConvLSTM \cite{ordonez2016deep} & 0.672 \\ \hline
LSTM-S \cite{hammerla2016deep} & 0.684 \\ \hline
Bi-LSTM \cite{hammerla2016deep} & 0.697 \\ \hline
Attention Model \cite{attention18} & 0.707 \\ \hline
LSTM Ensemble \cite{guan2017ensembles} & 0.726 $\pm$ 0.01 \\ \hline 
Attend$\&$Discriminate \cite{Attend_Discriminate} &0.746 \\ \hline \hline
Delay($\Delta=0.7s$)& 0.705 $\pm$ 0.02  \\ \hline
LSTM$\&$Inverse  & 0.733 $\pm$ 0.01  \\ \hline
LSTM(6)$\&$DLY(7)$\&$INV(7)  & \bf{0.778} $\pm$ 0.01 \\ \hline
\end{tabular}
\caption{\label{tab:opp_result} Model Comparison on OPP dataset.}
\end{table}


\begin{table}[htbp]
  \centering
    \begin{tabular}{||l|c|c||}
    \hline
    Model & Sensitivity & Mean F1 \bigstrut\\
    \hline
    CNN\cite{yang2015deep} & 0.804 & 0.658 \bigstrut\\
    \hline
    DeepConvLSTM\cite{ordonez2016deep}  & 0.823 & 0.665  \bigstrut\\
    \hline 
    
    Bi-LSTM\cite{hammerla2016deep}  & 0.814 &  0.682 \bigstrut\\
    \hline \hline
    Delay($\Delta=0.5s$)  & 0.865 $\pm$ 0.04  & 0.719 $\pm$ 0.01 \bigstrut\\
    \hline
    LSTM\&Inverse  & 0.833 $\pm$ 0.03 & 0.702 $\pm$ 0.01 \bigstrut\\
    \hline
    LSTM(6)$\&$DLY(7)$\&$INV(7)  & \bf{0.882} $\pm$ 0.03 & \bf{0.735} $\pm$ 0.01 \bigstrut\\
    \hline
    \end{tabular}%
\caption{Model Comparison on DG dataset (in both sensitivity and Mean-F1 score); Sensitivities correspond to a fixed specificity of 0.9.}
\label{tab:dg_result}
\end{table}%

\begin{table}[htbp]
  \centering
    \begin{tabular}{||c|c||}
    \hline
    Model & Mean F1 \bigstrut\\
    \hline
    CNN\cite{yang2015deep} & 0.771  \bigstrut\\
    \hline
    DeepConvLSTM\cite{ordonez2016deep}  & 0.765   \bigstrut\\
    \hline 
    Bi-LSTM\cite{hammerla2016deep}  & 0.802 \bigstrut\\
     
    \hline \hline
    Delay($\Delta=1.5s$) & 0.829 $\pm$ 0.04 \bigstrut\\
    \hline
    LSTM\&Inverse & 0.829 $\pm$ 0.05 \bigstrut\\
    \hline
    LSTM(6)$\&$DLY(7)$\&$INV(7)  & \bf{0.857} $\pm$ 0.03  \bigstrut\\
    \hline
    \end{tabular}%
  \label{tab:addlabel}%
    \caption{Model Comparison on PAMAP2 dataset}
    
\end{table}%

\subsection{Discussion}
In the tasks of recognizing the challenging sporadic activities, sample-wise LSTM demonstrated its great potential in previous works. 
In this work, we studied several training strategies (including the two LSTM variants) to further boost the performance for SAR, which can also generalize well to Non-SAR tasks.  

\subsubsection{LSTM variants as Simple Training Strategies}
Our delay and inverse models can be regarded as training strategies, which can be realized without changing the code of LSTM.
Specifically, the delay model can be trained by feeding the shifted signal sequences(i.e., backward by $\Delta$ timestamps, as shown in Fig.~\ref{fig:delay_model}), while the inverse model can be trained by feeding the inverted sequences, as shown in Fig.~\ref{fig:inverse_model}.
Without alteration of LSTM, these two variants are simple, and can be very practical. 

\subsubsection{LSTM$\&$Inverse and Bi-LSTM}
It is worth mentioning that our LSTM$\&$Inverse model and Bi-LSTM \cite{hammerla2016deep} are two different methods.
Although Bi-LSTM \cite{hammerla2016deep} can also leverage the backward sequential information, it is limited to the sliding-window level and tends to be short-term.
In contrast, inverse model can learn the long-term backward sequential information without the aforementioned limitation.
In essence, Bi-LSTM is to combine forward/backward LSTM units in the network, while LSTM$\&$Inverse is to perform score-level fusion of two separated LSTMs, which can exploit long-term (in both forward/backward) sequential information.

\subsubsection{Fusion Scheme for the General HAR}
Based on the epoch-wise bagging scheme (with HYPAV)\cite{guan2017ensembles}, much higher performance can be achieved if base learners are from multiple training sources (e.g., LSTM/Delay/Inverse models), than single source (e.g., LSTM ensemble \cite{guan2017ensembles}).
In the multi-source ensemble, higher diversity can be easily achieved since the fusion system is formed by base classifiers generated from different epoch-wise bagging schemes, which would also increase the training cost by a factor of source number. 
Nevertheless, for some challenging HAR cases this could be a potential way to boost the performance. 
It is worth noting the multi-source fusion can be very flexible, e.g., it can also be achieved by fusing epoch-wise classifiers from different network architectures.

\section{Conclusion}
In this work, we proposed two simple yet effective LSTM variants, namely delay and inverse models for tackling the challenging SAR problems.
They can be deemed as training strategies without alteration of the LSTM fundamentals.  
For time-critical SAR scenarios, the delay model can effectively take advantage of the delay interval for performance improvement.
For regular SAR tasks without time-critical requirement, the inverse model can be used to analyze the activities in an inverse manner, which is complementary to LSTM. 
Based on LSTM and its variants, we also studied additional training strategies, i.e., hyper-parameter as variable (HYPAV) and epoch-wise bagging \cite{guan2017ensembles}, and we found significant performance gains can be achieved when multi-source ensemble was applied in both SAR and non-SAR tasks, indicating their effectiveness in challenging HAR scenarios. 

\bibliographystyle{IEEEtran}
\bibliography{IEEEabrv,bibfile}

\end{document}